\documentclass[11pt,twoside]{article}
\usepackage[dvips]{graphics}
\usepackage{newpasp}
\newcommand{\oversim}[2]{\protect{\mbox{\lower0.5ex\vbox{%
   \baselineskip=0pt\lineskip=0.2ex
   \ialign{$\mathsurround=0pt #1\hfil##\hfil$\crcr#2\crcr\sim\crcr}}}}} 
\newcommand{\simgreat}{\mbox{$\,\mathrel{\mathpalette\oversim>}\,$}} 
\newcommand{\simless} {\mbox{$\,\mathrel{\mathpalette\oversim<}\,$}} 
\def\edcomment#1{\iffalse\marginpar{\raggedright\sl#1\/}\else\relax\fi}
\reversemarginpar

\begin{document}
\title{The Initial Mass Function and its Variation}
 \author{Pavel Kroupa} 

\affil{Institut f\"ur Theoretische Physik und Astrophysik\\ Universit\"at
Kiel, D-24098 Kiel, Germany}

\begin{abstract}
The observed distribution of IMF shapes can be understood as
statistical sampling from a universal IMF and variations that result
from stellar-dynamical processes. However, young star clusters appear
to have an IMF biased towards low-mass stars when compared to the
Galactic disk IMF, which comprises an average populated by
star-formation episodes occurring~Gyrs ago. In addition to this
tentative but exciting deduction, this text outlines some of the
stumbling blocks hindering the production of rigorous IMF
determinations, and discusses the most robust evidence for structure
in it. Also, it is stressed that an invariant IMF should not be
expected.
\end{abstract}

\keywords{IMF -- cluster dynamics -- mass segregation -- binary
systems -- massive stars -- brown dwarfs}

\section{Introduction}
\noindent
The stellar initial mass function (IMF) is one of the most fundamental
distribution functions in astrophysics, and a large effort is being
put into constraining its form and possible invariance.  Much progress
has been achieved since Salpeter published the first constraints in
1955. Most of the papers appearing in this field are observational in
nature, which is {\it necessary but not sufficient} for providing the
type of statistical sample necessary to address the two key questions,
namely (1) what is the {\it form of the IMF}, and (2) does this form
show {\it unambiguous systematic variations with star-forming
conditions}?

In addition to the significant observational effort, low-cost
modelling is crucial in distilling rigorous IMF properties from the
many observational results offered to the community. For example, one
young cluster may have a flatter IMF at low masses than another.  Can
such differences be attributed to different binary proportions, and if
so, can these be explained by the clusters being in different stages
of dynamical evolution, but {\it starting with the same initial
population}? Does the observed mass-ratio distribution of
Galactic-field binary systems reflect the initial conditions, or has
it been substantially modified through stellar-dynamical evolution in
young star clusters?  Can bumps, wiggles and depressions in the
observationally deduced IMF of a $\simless2$~Myr population be
attributed to incomplete stellar physics, such as applying idealised
non-rotating classical pre-main sequence contraction tracks of single
stars to a real ensemble of fast-rotating magnetically active multiple
stars that are still accreting and that retain a memory of their
accretion history?

Without addressing these kinds of questions, which mean significant
future theoretical stellar-dynamical and stellar-structure efforts in
studying the IMF, any statements made about its shape are of limited
scientific value. That this is a {\it necessary} complement to the
observational effort for advancing knowledge on the IMF is
demonstrated by the fact that apparent sub-structure in the IMF can be
explained by fine-structure in the mass--luminosity relation of stars
(Belikov et al. 1998; Kroupa 2001a), and that the Galactic-field
binary-star properties can be unified with those of pre-main sequence
stars if most stars form in embedded clusters, implying a {\it
surprising degree of universality of the primordial binary-star
properties as well as of the IMF} (Kroupa 1995; Kroupa, Aarseth \&
Hurley 2001).

After offering some technicalities in Section~\ref{sec:tech}, the
subject of the present text (see Kroupa 2001b for additional details)
is to discuss the statistical properties of the available data
(Section~\ref{sec:dets}), deduce possible evidence for a
systematically varying IMF (Section~\ref{sec:syst}), and to point to
particularly interesting questions that follow from the observational
data and the modelling thereof (Section~\ref{sec:quests}).

\section{The functional form and generating function}
\label{sec:tech}
\noindent
The number of stars in the mass interval $m$ to $m+dm$ is
$\xi(m)\,dm$.  Which particular functional form best approximates the
IMF, $\xi(m)$, is very much open to debate. The multi-part power-law
IMF has the merit that it is an {\it easy analytical tool} for
describing the IMF over broad mass-ranges, allowing controlled
variation of various parts of the IMF, such as what effects different
relative numbers of massive stars have on star-cluster evolution
without changing the form of the LF for low-mass stars.  Also, the
multi-part power-law IMF is easily transformed to the mass-generating
function, which allows efficient creation of theoretical stellar
populations.

Ensuring continuity leads to
\begin{equation}
\xi (m) = k\left\{
          \begin{array}{l@{\quad\quad,\quad}l}
   \left({m\over m_{\rm H}}\right)^{-\alpha_0}  &m_l < m \le m_{\rm H}, \\
   \left({m\over m_{\rm H}}\right)^{-\alpha_1}  &m_{\rm H} < m \le m_0, \\
   \left[\left({m_0\over m_{\rm H}}\right)^{-\alpha_1}\right] 
        \left({m\over m_0}\right)^{-\alpha_2} 
        &m_0 < m \le m_1,\\
   \left[\left({m_0\over m_{\rm H}}\right)^{-\alpha_1}
        \left({m_1\over m_0}\right)^{-\alpha_2}\right] 
        \left({m\over m_1}\right)^{-\alpha_3} 
        &m_1 < m \le m_2,\\
   \left[\left({m_0\over m_{\rm H}}\right)^{-\alpha_1}
        \left({m_1\over m_0}\right)^{-\alpha_2}
        \left({m_2\over m_1}\right)^{-\alpha_3}\right] 
        \left({m\over m_2}\right)^{-\alpha_4} 
        &m_2 < m \le m_u,\\
          \end{array}\right.
\label{eqn:imf_mult}
\end{equation}
where $k$ contains the desired scaling. Note that the present data
only support a three-part power-law IMF (eqn~\ref{eq:imf} below).

If $N_{\rm tot}$ is the total number of stars in some star-formation
event, then the IMF can be interpreted as a probability density,
$p(m)=\xi(m)/N_{\rm tot}$.  The probability of picking a stellar mass
in the range $m$ to $m+dm$ becomes $dX=p(m)dm$, so that
$\int_{m_l}^{m_u}p(m')dm' = 1; X_{m_l}^m = \int_{m_l}^{m}p(m')dm'$.
From this the generating function, $m(X)$, is derived, with $X$ a
discrete random deviate distributed uniformly between~0 and~1.  The
normalisation condition gives $X_l^{\rm H}, X_{\rm H}^0, X_0^1, X_1^2,
X_2^u$ with $1=X_l^{\rm H} + X_{\rm H}^0 + X_0^1 + X_1^2 + X_2^u$. For
example, if $X$ falls into the range $X_l^{\rm H} + X_{\rm H}^0 < X
\le X_l^{\rm H} + X_{\rm H}^0 + X_0^1$ (i.e. $m_0< m \le m_1$), then
the corresponding stellar mass is generated from $m(X) =
\left[m_0^{1-\alpha_2} + \left(X-X_l^{\rm H}-X_{\rm H}^0\right)
{\left(1-\alpha_2\right)\,K \over \beta_2}\right]^{1\over 1-\alpha_2},
$ where $\beta_2=m_0^{\alpha_2}\left(m_{\rm
H}/m_0\right)^{\alpha_{1}}$ from eqn.~\ref{eqn:imf_mult}, and
$K=N_{\rm tot}/k$.

\section{The empirical alpha-plot}
\label{sec:dets}
\begin{figure}
\plotfiddle{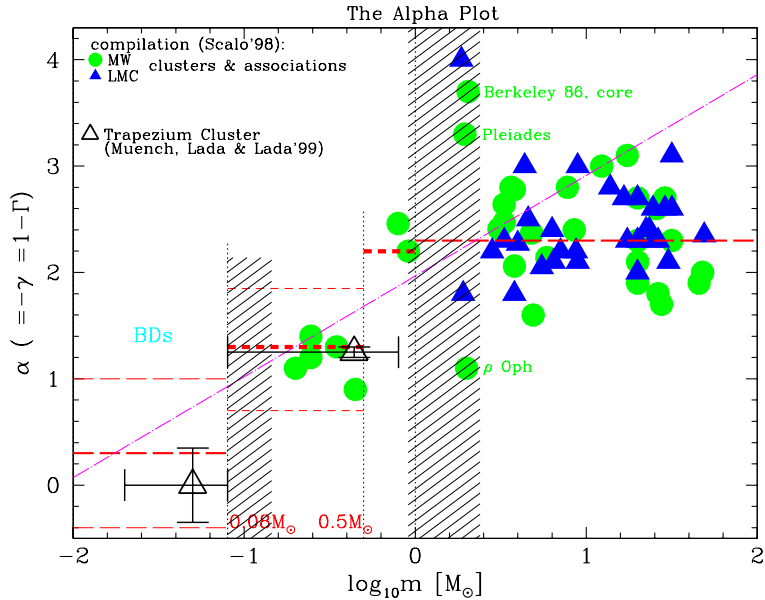}{7cm}{-90}{40}{40}{-165}{230}
\caption{\small The {\it alpha-plot} for Milky-Way (MW) and
Large-Magellanic-Cloud (LMC) star-clusters and associations. Data are
from the compilation of Scalo (1998).  The horizontal dashed lines are
the average IMF with estimated uncertainties (eqn~\ref{eq:imf}), and
the diagonal dot-dashed line is the popular Miller-Scalo (1979) IMF.
From Kroupa (2001b).  The figure can be obtained in colour from
astro-ph/0009005.}
\label{fig:apl}
\end{figure}

\noindent
An exciting way to summarise the many IMF estimates is the alpha-plot
(Fig.~\ref{fig:apl}) which displays the determinations of the IMF
power-law indices, $\alpha$, over the average logarithm of the mass
interval over which $\alpha$ is derived.
Fig.~\ref{fig:apl} implies (additional details can be found in
Kroupa 2001b):

\vspace{-3.5mm}

\begin{enumerate}
\itemsep=-5mm
\item
The MW and LMC data show no systematic difference despite different
average metallicities. $\alpha$ is thus at most weakly dependent on
[Fe/H] for $m\simgreat 3\,M_\odot$. \\

\item
The IMF steepens progressively with increasing mass. \\

\item
$\alpha$ appears to become constant near the Salpeter value for
$m\simgreat 0.5\,M_\odot$. \\

\item
For $0.1-1\,M_\odot$ the cluster data are beautifully consistent with
the Galactic-field IMF obtained from detailed star-count analysis
using different mass-luminosity relations {\it and corrected for
unresolved binaries}. Note that the Galactic-field IMF is especially
tightly constrained for $0.5\simless m\simless 1\,M_\odot$. \\

\item 
The scatter in $\alpha$ is small for $m \simless 0.8\,M_\odot$.\\

\item
The scatter is large but constant and apparently non-Gaussian for
$m\simgreat 3\,M_\odot$. \\

\item
The shaded regions indicate mass-ranges that are particularly prone to
errors.  At the low-mass end the long contraction times to the main
sequence make mass estimates very difficult when the ages are
unknown. Near $2\,M_\odot$ evolution along and off the main sequence
becomes problematical for the same reason, and stars in this mass
range are typically the most massive in the cluster samples entering
Fig.~\ref{fig:apl}, and thus especially prone to stellar-dynamical
biases (mass segregation and ejections). Not much emphasis is thus
placed on the large observational scatter in the shaded region near
$2\,M_\odot$.\\
\end{enumerate}

\vspace{-6.5mm}

\noindent 
Based on the above and previous work, Kroupa (2001b) approximates the
{\it average Galactic-field IMF} with
\begin{equation}
          \begin{array}{l@{\quad\quad,\quad}l}
\alpha_0 = +0.3\pm0.7   &0.01 \le m/M_\odot < 0.08, \\
\alpha_1 = +1.3\pm0.5   &0.08 \le m/M_\odot < 0.50, \\
\alpha_2 = +2.3\pm0.3   &0.50 \le m/M_\odot, \\
          \end{array}
\label{eq:imf}
\end{equation}
($\alpha_4=\alpha_3=\alpha_2$, the uncertainties reflect 95--99~per
cent confidence intervals).  This is an average because the
Galactic-field is composed of many star-formation events that occurred
Gyrs ago. Note that Salpeter (1955) estimated $\alpha=2.35$ for stars
in the mass interval $0.4-10\,M_\odot$. Any confident variation of the
IMF should be evident in {\it significant} departures from
eqn~\ref{eq:imf}.

\section{A systematically varying IMF}
\label{sec:syst}
\noindent
The MW clusters plotted in Fig.~\ref{fig:apl} are between a few and
100~Myr old, and thus sample the latest episode of star-formation in
the MW disc near the solar circle. The immediate but naive conclusion
follows that the IMF remained the same over the age of the Galactic
disc, at least to within the uncertainties apparent in
Fig.~\ref{fig:apl}. Perhaps this would not be surprising if most stars
form in clusters (Kroupa 1995).  However, what is the origin of the
scatter of $\alpha$ values?

\subsection{The theoretical alpha-plot}
\label{sec:theo}
\noindent
To address this problem, a large library of stellar-dynamical models
was evolved for 150~Myr using a modified version of Aarseth's (1999)
{\sc Nbody6} code (costing a few months of CPU time on standard
desk-top computers) of binary-rich clusters that have initially the
same large central density (and thus the same crossing time) as the
Orion Nebula Cluster, but different number of stars ($N=800, 3000,
10^4$). Stellar evolution is included and the initial
binary-properties are consistent with pre-main sequence
constraints. Stars are initially paired randomly from the IMF
(eqn~\ref{eq:imf}), but the mass-ratio distribution is rapidly
depleted of low-mass companions.  The MFs in the inner and outer
regions of the clusters are constructed at 3~Myr and 70~Myr counting
the masses of systems (single stars and remaining binaries) giving the
system MF. Theoretical alpha-plots are generated to study the scatter
in $\alpha$ values.

\begin{figure}
\plotfiddle{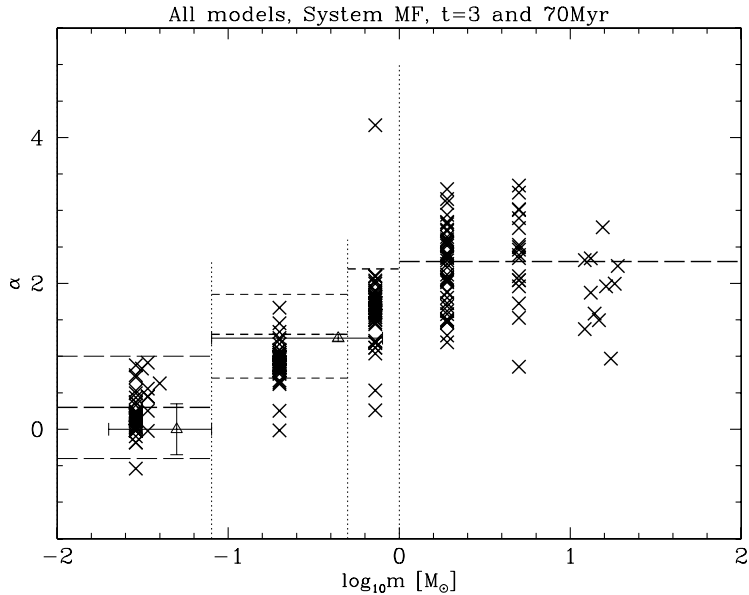}{6.8cm}{-90}{40}{40}{-165}{225}
\caption{\small The theoretical alpha-plot. From Kroupa (2001b).
$\alpha$ is measured using least-squares fitting to segments of the MF
that have the average log-masses indicated by the discrete values.}
\label{fig:ath}
\end{figure}

The resulting data are combined in Fig.~\ref{fig:ath}. It is evident
that the scatter is reproduced, it being approximately constant for
$m\simgreat1\,M_\odot$. The unsettling finding, however, is that the
system MFs show systematically smaller mean $\alpha$ for
$0.1<m<1\,M_\odot$ than the input IMF (eqn~\ref{eq:imf}). The bias
amounts to $\Delta\alpha\approx0.5$.

The binary proportion, $f$, decreases on the same time-scale in all
clusters because they have the same crossing time. By a few~Myr the
binary population stabilises near $f=0.4-0.5$ depending on the initial
number of stars in the cluster. For BDs $f\approx0.2$ owing to their
small binding energy.  Real clusters are likely to have a larger $f$
(e.g. K\"ahler finds that $f\approx0.7$ is possible for the Pleiades
cluster), which would not be surprising given that the present models
start from a very dense initial state leading to efficient binary
disruption. Thus, if anything, {\it the bias in $\alpha$ is likely to
be underestimated with the present calculations}.

\subsection{Implications}
\label{sec:impl}
\noindent
The cluster and association data plotted in Fig.~\ref{fig:apl} are not
corrected for unseen companion stars.  The implication is that these
data must be systematically larger by at least $\Delta\alpha=0.5$ for
low-mass stars. The bias is nearly negligible for BDs because they
develop a small binary proportion and most BDs in stellar systems are
disrupted from their stellar primaries.  Thus, {\it present-day
star-formation seems to be producing more low-mass stars today than
when most of the Galactic disk formed}.  The revised, present-day
star-formation IMF thus reads
\begin{equation}
          \begin{array}{l@{\quad\quad,\quad}l}
\alpha_0 = +0.3\pm0.7   &0.01 \le m/M_\odot < 0.08, \\
\alpha_1 = +1.8\pm0.5   &0.08 \le m/M_\odot < 0.50, \\
\alpha_2 = +2.7\pm0.3   &0.50 \le m/M_\odot < 1.00, \\
\alpha_3 = +2.3\pm0.3   &1.00 \le m/M_\odot, \\
          \end{array}
\label{eq:revimf}
\end{equation}
with $\alpha_4=\alpha_3$.

\vfill\newpage

\section{Important questions}
\label{sec:quests}
\noindent

\vspace{-5mm}

\begin{enumerate}
\itemsep=-4.5mm

\item Is it true, that the Galactic disk is now producing more
low-mass stars than at an earlier epoch (eqn~\ref{eq:revimf})?  The
most obvious physical parameter that changes with time is the overall
metallicity and thus rate of cooling of a molecular cloud.
Theoretical arguments lead to the expectation that such a variation
ought to be the case, although it's magnitude is uncertain (Larson
1998).

For this systematic effect to go away, the solar-neighbourhood IMF
would need to be steeper than most available studies imply (Kroupa
2001a).\\

\item Fig.~\ref{fig:ath} shows that for $m\simgreat6\,M_\odot$ the
theoretical $\alpha(m)$ data fall towards smaller values. This results
from under-sampling near the maximum stellar mass, and may be an
artefact of the present models. However, observational data probably
also suffer from such an effect, the possible implication being that
the true underlying IMF may {\it steepen for
$m\simgreat6\,M_\odot$}. \\

\item The modelling by Sagar \& Richtler (1991) demonstrates that for
a high binary proportion among massive stars (massive primaries {\it
and} secondaries), the IMF is steeper by as much as $\Delta\alpha=0.4$
or more. It is thus possible, that for $m \simgreat 1\,M_\odot$,
$\alpha\simgreat2.7$, keeping in mind the previous point.\\

\item What is the Galactic-field IMF for $m\simgreat1\,M_\odot$?  {\it
We have no direct information on this}! We only have constraints on
the IMF for intermediate and massive stars for present-day
star-formation in clusters and OB associations, and eqn~\ref{eq:imf}
is pure speculation for $m>1\,M_\odot$ (the dichotomy problem stated
in Kroupa 2001b).\\

\item So far not discussed, but very important, is the point made by a
number of authors (e.g. Eisenhower 2001) that some star-bursts
indicate IMFs that are top-heavy. How reliable is this evidence? This
cannot be answered until high-precision stellar-dynamical models of
realistic very young massive star-clusters that typically form en
masse in star-bursts have become available, and the reliability of the
stellar models for young massive stars are improved. Thus, for
example, a kinematically decoupled core of massive stars may be
embedded in a cluster that predominantly consists of low-mass
stars. The core can have a velocity dispersion typical for the massive
sub-population, but not indicative of the entire cluster. Available
stellar-dynamical models of moderately massive clusters do indicate
that mass segregation is very rapid, on the order of a few initial
crossing times, so that by a few Myrs a decoupled core of OB stars may
have formed in most massive clusters (Kroupa 2001c). Until
stellar-dynamical libraries, such as is being constructed by Kroupa
(2001c), are not analysed in this respect taking into account the
observations, the statements on IMFs in star bursts remain all too
speculative.\\

\item What is the physical basis for the structure evident in the IMF
(flattening near 0.5 and $0.08\,M_\odot$, eqn~\ref{eq:imf})? Adams \&
Fatuzzo (1996), Padoan \& Nordlund (2001), Klessen (2001b) and Bonnell
et al. (2001) attempt various explanations.  \\

\end{enumerate}

\vfill\newpage

\section{Concluding remarks}
\label{sec:rem}
The theoretical data presented in Fig.~\ref{fig:ath} imply that
the scatter in the alpha-plot can be understood as being stochastic in
nature (as also stressed by Elme\-green 1999), together with deviations
caused by stellar-dynamical processes. It is surprising though that
the observational data in Fig.~\ref{fig:apl} show a comparable
distribution to the theoretical data plotted in Fig.~\ref{fig:ath},
indicating that there are no additional stochastic effects and that
the IMF is close to being invariant.

The bias through unresolved binary systems, however, implies that for
low-mass stars the cluster-data require a present-day IMF that is {\it
systematically} steeper than the solar-neighbourhood IMF. The argument
leading to this deduction is simple: An average IMF is calculated from
the cluster and association data in Fig.~\ref{fig:apl} giving
eqn~\ref{eq:imf} (ignoring the Galactic-field constraints). This
average IMF is used to construct theoretical alpha-plots of realistic
cluster populations, leading to contradiction with the empirical data
that are the basis for the input IMF.  The systematic disagreement
between the theoretical and empirical data imply that the true
underlying IMF must be steeper for low-mass stars
(eqn~\ref{eq:revimf}) than the assumed IMF (eqn~\ref{eq:imf}), the
latter also being the Galactic-field IMF.

This result opposes the cherished belief that the IMF is
universal. But this conviction is not intuitive anyway, since the
fundamental physics of star formation provides rather convincing
arguments for variations. If the systematic variation is true, then it
must be even more evident in extreme populations. Thus, metal-poorer
populations ought to have an IMF with systematically smaller $\alpha$
for low-mass stars. While observations tentatively suggest this may be
the case for globular clusters, the evidence is ambiguous, because
dynamical evolution leads to the preferential loss of low-mass
stars. The white-dwarf (WD) candidates in the Galactic halo may
represent a very early phase of star-formation in the Galaxy, and the
absence of associated low-mass stars that ought to still burn H
suggests again the same sense of systematic evolution of the
IMF. However, this evidence too is not yet rigorous, since the nature
of the WD candidates must first be illuminated (Chabrier 1999).

Even within an embedded star-cluster systematic variations ought to
occur, because near the high-density core of the cluster proto-stars
interact if the proto-star collapse time-scale is comparable to or
longer than the local cluster crossing time-scale.  Thus, massive
stars are expected to form near cluster centres (Bonnell, Bate \&
Zinnecker 1998; Klessen 2001a; Bonnell et al. 2001), which is also
suggested by observation (e.g. Hillenbrand 1997) and the requirement
of producing enough run-away OB stars (Kroupa 2001c). Once a massive
star ignites, further star-formation is compromised through the
explosively propagating HII region, which may affect the number of BDs
and less massive objects, at least some of these possibly being
unfinished stars. The sub-stellar part of the IMF should thus also
show variations depending on environment, and a tentative hint at this
has been reported (Luhman 2000). However, again it must be stressed
that this evidence may be elusive because the stellar-dynamical models
also lead to an apparent overabundance of BDs in dense regions simply
because such clusters are dynamically evolved leading to many more
star--BD and BD--BD systems having been disrupted (more BDs for the
observer to see).

\vfill\newpage




\begin{references}
{\small

\reference Aarseth, S. J. 1999, \pasp, 111, 1333

\reference Adams, F.\ C., Fatuzzo, M.\ 1996, \apj, 464, 256 

\reference Belikov, A.\ N., Hirte, S., Meusinger, H., Piskunov, A.\
	E.\ \& Schilbach, E.\ 1998, \aap, 332, 575

\reference Bonnell, I. A., Bate, M. R., Zinnecker, H. 1998, \mnras, 298, 93

\reference Bonnell, I. A., Clarke, C. J., Bate, M. R., Pringle, J. E.
	2001, \mnras, in press (astro-ph/0102121)

\reference Chabrier G., 1999, \apj, 513, L103

\reference Eisenhauer, F. 2001, in Starbursts: Near and Far,
     eds. L.J. Tacconi and D. Lutz, in press (astro-ph/0101321)

\reference Elmegreen, B. G. 1999, \apj, 515, 323

\reference Hillenbrand, L. A. 1997, \aj, 113, 1733

\reference K\"ahler, H. 1999, \aap, 346, 67

\reference Klessen, R. 2001a, ApJL, in press (astro-ph/0101277)

\reference Klessen, R. 2001b, ApJ, submitted

\reference Kroupa, P. 1995, \mnras, 277, 1507

\reference Kroupa, P. 2001a, in ASP Conf. Ser., Star2000: 
	Dynamics of Star Clusters and the Milky Way, eds. 
	S. Deiters, R. Spurzem, et al., in press (astro-ph/0011328)

\reference Kroupa, P. 2001b, \mnras, in press (astro-ph/0009005)

\reference Kroupa, P. 2001c, in ASP Conf. Ser., IAU Symp.200: The
	Formation of Binary Stars, eds. H. Zinnecker, R. Mathieu, in
	press (astro-ph/0010347)

\reference Kroupa, P., Aarseth, S. J., Hurley, J. 2001, \mnras, in
	press (astro-ph/0009470)

\reference Larson, R. B. 1998, \mnras, 301, 569

\reference Luhman, K. 2000, \apj, 544, 1044

\reference Miller, G. E., Scalo, J. M. 1979, ApJS, 41, 513

\reference Muench, A. A., Lada, E. A., Lada, C. J. 2000, \apj, 533, 358

\reference Padoan, P., Nordlund, A. 2001, \apj, submitted
	(astro-ph/0011465)

\reference Sagar, R., Richtler, T. 1991, \aap, 250, 324

\reference Salpeter, E. E. 1955, \apj, 121, 161

\reference Scalo, J. M. 1998, in ASP Conf. Ser. Vol. 142, The Stellar
           Initial Mass Function, eds. G. Gilmore \& D. Howell (San
           Francisco: ASP), p.201


}
\end{references}
\end{document}